\documentclass[journal=jacsat,manuscript=article]{achemso}
\usepackage[version=3]{mhchem}
\usepackage{amsmath,graphicx,color,verbatim}
\usepackage{graphicx}
\usepackage{xspace}
\usepackage{longtable}
\usepackage{booktabs}
\def\beq{\begin{equation}}
\def\eeq{\end{equation}}

\usepackage{adjustbox}
\usepackage{graphicx}
\usepackage{rotating}
\usepackage{longtable}
\usepackage{xr}
\usepackage{hyperref}
\SectionNumbersOn

\author{Manish Kothakonda}
\affiliation{Department of Physics and Engineering Physics, Tulane University, New Orleans, LA 70118}

\author{Ruiqi Zhang}
\affiliation{Department of Physics and Engineering Physics, Tulane University, New Orleans, LA 70118}

\author{Jinliang Ning}

\affiliation{Department of Physics and Engineering Physics, Tulane University, New Orleans, LA 70118}
\author{James Furness}

\affiliation{Department of Physics and Engineering Physics, Tulane University, New Orleans, LA 70118}

\author{Abhirup Patra}
\affiliation{Delaware Energy Institute, University of Delaware, 221 Academy Street, Newark, Delaware 19716, United States\\
(Presently at Shell Technology Center, Houston, TX 77082, United States)}

\author{Qing Zhao}
\affiliation{Department of Chemical Engineering, Northeastern University, Boston, MA 02115
}

\author{Jianwei Sun}
\email{jsun@tulane.edu}

\affiliation{Department of Physics and Engineering Physics, Tulane University, New Orleans, LA 70118}

\title{Towards chemical accuracy for chemi- and physisorption with an efficient density functional}

\begin{document}

\begin{tocentry}
\centering
\includegraphics[]{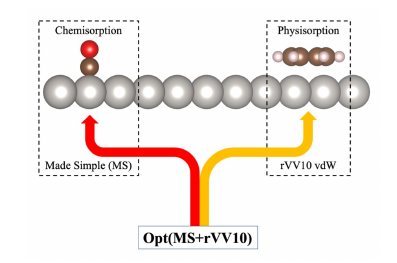}
\end{tocentry}

\begin{abstract}
Understanding molecular adsorption on surfaces underpins many problems in chemistry and materials science. Accurately and efficiently describing the adsorption has been a challenging task for first-principles methods as the process can involve both short-range chemical bond formations and long-range physical interactions, e.g., the van der Waals (vdW) interaction. Density functional theory presents an appealing choice for modeling adsorption reactions, though calculations with many exchange correlation density functional approximations struggle to accurately describe both chemical and physical molecular adsorptions. Here, we propose an efficient density functional approximation that is accurate for both chemical and physical adsorption by concurrently optimizing its semilocal component and the long-range vdW correction against the prototypical adsorption CO/Pt(111) and Ar$_2$ binding energy curve. The resulting functional opens the door to accurate and efficient modeling of general molecular adsorption.

\end{abstract}

% \date{\today}
\maketitle

\section{Introduction}

Molecular adsorption on solid surfaces is a challenging yet critical topic in surface science as a fundamental reaction step for surface chemical reactions. Understanding this processes of molecules adsorbing to solid surfaces is an essential step in understanding important surface phenomena including semiconductor processing, corrosion, electrochemistry, and heterogeneous catalysis\cite{sit2021density}. In catalysis, adsorption energies are particularly significant because they directly influence reaction rates, selectivity, and the stability of intermediates, all of which are essential for efficient catalyst design. Predicting adsorption energies accurately is vital for optimizing catalytic activity, as they determine stable intermediates, accessible transition states, and, ultimately, guide the design of active, selective, and durable catalytic materials. Misestimations in adsorption energy can lead to inaccurate predictions, hindering the development of improved catalysts essential for industrial and sustainable applications.

However, accurately understanding molecular adsorption on solid surfaces is challenging due to the limitations of available first-principles methods in describing interactions between finite molecules and semi-infinite solid surfaces\cite{swenson2019langmuir}. This process can involve the formation of both short-range chemical bonds, termed chemisorption, and long-range physical interactions, e.g., the van der Waals (vdW) interaction, termed physisorption. Calculations using high-level methods such as quantum Monte Carlo (QMC)\cite{foulkes2001quantum}, coupled-cluster singles
and doubles with perturbative triples CCSD(T)\cite{raghavachari1989fifth}, and random-phase approximation (RPA)\cite{hesselmann2017low,schimka2010accurate} can capture both chemisorption and physisorption accurately but their severe computational cost prevents their routine application to molecular adsorption problems. Calculations using density functional theory (DFT) \cite{hohenberg1964inhomogeneous,kohn1965self} are comparatively cheap however, this combination of computational efficiency and useful accuracy has result
ing in their becoming a widely accepted workhorse method for surface calculations. The past decades in particular have seen extraordinary progress in DFT driven computational heterogeneous catalysis \cite{norskov2004origin, norskov2011density}, resulting in greatly accelerated computer based catalyst design \cite{goldsmith2018machine,du2014catalyst,hammer2000theoretical}. Nevertheless, the limitations of exchange-correlation density functional approximations (DFAs) prevent DFT from reliably predicting both chemisorption and physisorption interactions with equal accuracy, thus posing a significant challenge in catalytic material design.

The accuracy of a DFT calculation is largely determined by the chosen exchange-correlation approximation, very many of which have been proposed. Approximate exchange-correlation functionals can broadly be categorized into a hierarchy of increasing sophistication accompanied by increasing computational cost and (ideally) increasing accuracy \cite{Jacobsladder}. The generalized gradient approximation (GGA) and meta-GGA class of exchange-correlation functionals are well suited for computing surface quantities in chemistry and condensed matter physics since these approximations require only semilocal ingredients and hence remain computationally inexpensive \cite{zhang2018efficient,pokharel2022sensitivity}. The absence of non-local ingredients means that long-range van der Waals interactions are absent from GGAs and meta-GGAs however, which can add significant errors to surface calculations \cite{rodriguez2014van,su2019modeling,jones1989density, patra2017properties}. To remedy this, various vdW corrections have been developed to pair with the semilocal approximations \cite{grimme2010consistent, grimme2006semiempirical,D3,D4,dion2004van, lee2010higher, cooper2010van, klimevs2009chemical, klimevs2011van, hamada2014van, berland2014exchange, tkatchenko2009accurate,ruiz2016density}. One such case, highly relevant to this work, involved pairing the meta-GGA MS2\cite{sun2012communication,sun2013semilocal} with the rVV10\cite{rVV10} vdW correction without optimizing its parameters. This resulted in a functional that failed to accurately capture the vdW wells, particularly for the well-studied H$_2$ + Cu(111) system\cite{smeets2021performance}. Typically, the vdW correction is fitted independently while the GGA or meta-GGA functional retains its parameterization based on general chemistry applications. This approach often leads to a trade-off: most GGA/meta-GGA functionals with vdW corrections perform well for either chemisorption or physisorption, but not both simultaneously. Because the semilocal functional and vdW correction are not designed in conjunction, existing GGA and meta-GGA functionals, even with vdW corrections, struggle to capture both types of adsorption with high accuracy at the same time. {\color{blue}Recent advances in many-body dispersion methods, particularly the work by Tkatchenko and colleagues on screened van der Waals interactions and many-body dispersion (MBD) effects\cite{tkatchenko2009accurate,ruiz2012density,maurer2016adsorption}, have demonstrated the importance of collective electronic effects in surface adsorption beyond pairwise additivity.
}

In this work we propose a novel density functional  with balanced performance for both chemi- and physisorption by concurrently optimizing both semilocal component and the long-range vdW correction against the prototypical CO/Pt(111) and the Ar$_2$ binding energy curves, see Section \ref{sec_params} for further details. The resulting density functional, termed ``Opt(MS+rVV10)'' which is the topic of this work is a reparameterization of both the meta-GGA Made Simple (MS) exchange-correlation functional \cite{sun2012communication} and the rVV10 long range vdW correction \cite{rVV10}. Opt(MS+rVV10) shows improved and balanced performance for both the physisorption and chemisorption of molecules adsorbed on transition metal surfaces compared to the other DFAs popular for surface science \cite{wellendorff2015benchmark}. Furthermore, Opt(MS+rVV10) shows the most balanced performance for molecular adsorptions of the DFAs considered and yields both the chemi- and physisorption local minima for graphene adsorbed on a Ni(111) surface, predicting a binding energy curve in close agreement with that from the high-level RPA \cite{olsen2011dispersive}.

\section{Results}
\subsection{Bivariate Plot}
\begin{figure*}
\centering
{\includegraphics[width=\textwidth]{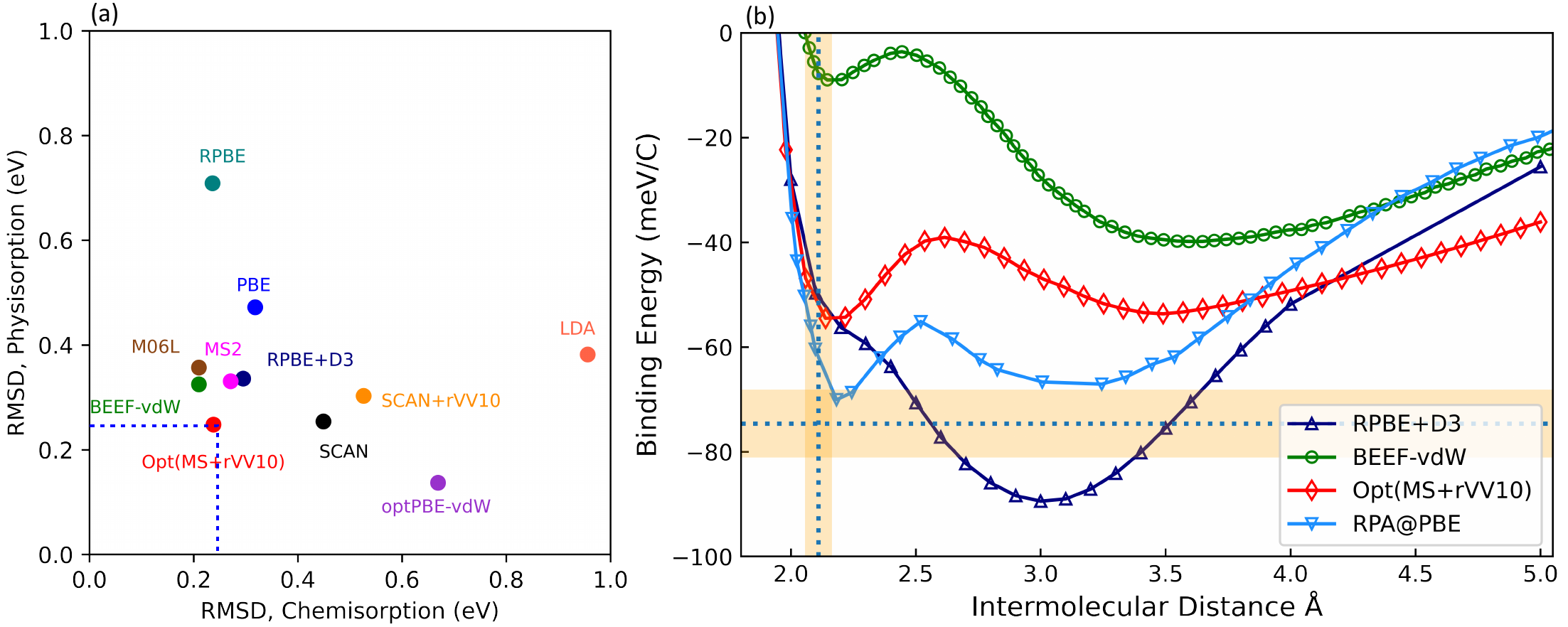}}
\caption{(a) Bivariate plot of Root Mean Square Deviation (RMSD) of the chemisorption and physisorption energies per adsorbate for CE39 systems. The blue dashed line represents the 0.25 eV RMSD threshold for both physisorption and chemisorption, with only Opt(MS+rVV10) edging this boundary. (b) Binding energy curves for graphene adsorbed in Ni(111) surface from RPBE+D3\cite{hammer1999improved}, BEEF-vdW \cite{shepard2019first,wellendorff2012density}, and Opt(MS+rVV10) compared with RPA\cite{olsen2011dispersive} results. The blue dotted lines indicate the experimental adsorption distance \cite{gamo1997atomic} of $2.11 \pm 0.07$~\r{A}, representing the uncertainty range in the measurement, while the orange shaded regions correspond to the experimental adsorption energy range \cite{shelton1974equilibrium, janthon2013theoretical} of $74 \pm 8.1$~meV, reflecting the uncertainty in the reported values.}
\label{fig:Bivatiate_graphene_Ni}
\end{figure*}

The CE39 dataset, proposed by Wellendorf et al. \cite{wellendorff2015benchmark}, comprises experimental data for 39 well-defined adsorption energies of various gas molecules adsorbed on eight transition metal surfaces. {\color{blue}These experimental adsorption energies have been corrected for zero-point vibrational energy (ZPVE) and RT contributions, following the procedure outlined in Ref.\citenum{wellendorff2015benchmark}, to approximate 0 K enthalpies. This correction enables a consistent and meaningful comparison to static-lattice DFT energies, which are calculated at 0 K without thermal contributions.} It includes two subsets: strongly bound chemisorbed systems and weakly bound physisorbed systems where van der Waals (vdW) interactions dominate. Figure \ref{fig:Bivatiate_graphene_Ni}(a) presents a bivariate plot contrasting the performance of Opt(MS+rVV10) against other widely used density functionals, such as the local density approximation (LDA\cite{parr1980density}), generalized gradient approximations (PBE\cite{PBE} and RPBE\cite{hammer1999improved}), meta-GGAs (SCAN\cite{SCAN}, MS2\cite{MS2_2}, and M06L), vdW-corrected GGAs (RPBE+D3, BEEF-vdW\cite{wellendorff2012density}, and optPBE-vdW), and vdW-corrected meta-GGAs (SCAN+rVV10). These functionals were selected due to their popularity in surface chemistry (e.g., RPBE) and broader applications (e.g., PBE).
The LDA functional significantly overestimates chemisorption with an root mean square deviation (RMSD) around 1.0 eV but predicts physisorption with reasonable accuracy (0.4 eV RMSD). This unexpected accuracy in physisorption likely results from error cancellation, as LDA generally overestimates electron density overlap bonding while failing to account for long-range vdW interactions.PBE, a GGA functional, improves upon LDA for chemisorption, reducing the RMSD from 1.0 eV to 0.38 eV, while RPBE further lowers it to 0.22 eV. However, both functionals struggle with physisorption, exhibiting RMSDs of 0.46 eV and 0.71 eV for PBE and RPBE, respectively. More recently, Sharada et al.\citenum{mallikarjun2019adsorption} extended the CE39 set by adding two additional adsorption reactions, creating the AD41 benchmark, which has been used to assess new density functional approximations such as the empirically fitted meta-GGA MCML\cite{brown2021mcml}. This functional, while grounded in physical constraints and informed by experimental and quantum chemistry reference data, is empirically optimized to yield improved accuracy for surface and gas-phase reaction energetics.
% PBE, a GGA functional, improves upon LDA for chemisorption (0.38 eV RMSD), while RPBE offers further improvement (0.22 eV RMSD). 
% However, both functionals exhibit notable errors in physisorption, with RMSDs of 0.46 eV for PBE and 0.71 eV for RPBE.

When vdW corrections are applied, RPBE+D3 \cite{ImprovedRPBED3} significantly enhances RPBE's performance, delivering a more balanced description of both chemisorption (0.31 eV RMSD) and physisorption (0.35 eV RMSD). The BEEF-vdW functional, developed using a Bayesian approach to fit exchange-correlation parameters for surface chemistry \cite{wellendorff2012density}, achieves an even better RMSD for chemisorption (0.21 eV) compared to RPBE+D3 and a slightly improved physisorption RMSD (0.34 eV). However, this imbalance may be attributed to the fact that 17 chemisorbed systems from the CE39 dataset were included in the BEEF-vdW training set. While optPBE-vdW performs best for physisorbed systems among the tested functionals, its overestimation of chemisorption remains unsatisfactory \cite{hensley2017dft}.

At the meta-GGA level, although SCAN \cite{SCAN} has been highly successful in addressing longstanding issues in condensed matter physics, such as strongly correlated cuprates and phase transitions \cite{furness2018accurate,zhang2017comparative, pokharel2022sensitivity}, it overestimates chemisorption with an RMSD of approximately 0.44 eV. This may be due to self-interaction errors, which can lead to an overestimation of charge transfer between molecules and metal surfaces, as evidenced in CO/Pt(111) adsorption studies \cite{patra2019rethinking}. SCAN, however, has demonstrated an ability to capture intermediate vdW interactions \cite{sun2016accurate}, making it reasonably accurate for physisorption with an RMSD of around 0.30 eV. Another meta-GGA, M06L, performs exceptionally well for chemisorption, with an RMSD of 0.21 eV, but it is significantly less accurate for physisorption (RMSD of 0.37 eV). This is surprising, given that M06L was the first semilocal functional to successfully capture intermediate vdW interactions \cite{madsen2010treatment}. The discrepancy may be due to the highly empirical nature of M06L, which over-represents molecular systems, leading to an exaggerated emphasis on chemical bonding at the expense of accurately modeling physisorption.

The addition of the rVV10 vdW correction to SCAN (SCAN+rVV10) results in an overestimation of chemisorption relative to SCAN, while also degrading SCAN’s performance in physisorption. This underscores the delicate balance required when pairing a semilocal density functional with a long-range vdW correction.

Even without a vdW correction, the MS2 meta-GGA remains one of the most balanced functionals, offering reasonable accuracy across both chemisorption and physisorption. This balance motivated the pairing of MS2 with the rVV10 vdW correction, alongside a re-optimization of the internal parameters, to enhance accuracy while maintaining balanced performance. As demonstrated, the resulting Opt(MS+rVV10) functional provides the best balanced performance for both chemisorption (RMSD$\sim$ 0.24 eV) and physisorption (RMSD$\sim$ 0.26 eV).

\subsection{Graphene adsorption on Ni(111)}

As the CO/Pt(111) and Ar$_2$ systems were used to optimize the parameters, we test the transferability of Opt(MS+rVV10) for chemisorption and physisorption by applying it to graphene adsorbed on Ni(111). This system was not part of parameterization and is believed to have a challenging double minima of both chemisorption and physisorption in its binding energy curve.

Graphene adsorption on metal is highly metal dependent forming strong chemical bonding with some metals and only weak van der Waals interaction with others. Graphene on Ni has shown exceptional electronic properties in semiconducting technology \cite{dai2011rational}, however understanding the interactions between graphene and Ni(111) surface has been challenging for both experiment and theory. A study using angle resolved photoemission (ARPES) reveals strong chemical bonding combined with weak vdW interactions between the graphene sheet and Ni(111) \cite{gruneis2008tunable}. Several theoretical studies have been made drawing conflicting conclusions; some predicting dominant chemisorption minima and others predicting deeper physisorption minima\cite{bertoni2005first,kalibaeva2006ab,giovannetti2008doping,tao2018modeling,vanin2010graphene}. High level RPA calculations have established that both chemi- and phsysisorption minima should have similar depth, in good accord with the experimentally determined binding energy, as shown in Figure \ref{fig:Bivatiate_graphene_Ni}(b). Opt(MS+rVV10) predicts chemisorption and physisorption minima at 2.15 \r{A} and 3.5 \r{A}, respectively, compared to RPA@PBE values of 2.17 \r{A} and 3.27 \r{A}. The binding energies for these minima are -58 meV and -57 meV for Opt(MS+rVV10), versus -70 meV and -67 meV for RPA@PBE.

The binding energy curves of graphene adsorbed on Ni surface calculated using: RPBE+D3, BEEF-vdW, and Opt(MS+rVV10), are shown in Figure \ref{fig:Bivatiate_graphene_Ni}(b). These functionals were chosen as representative vdW corrected functionals that exhibit good balance between chemi- and physisorption in Figure \ref{fig:Bivatiate_graphene_Ni}(a). The binding curves for BEEF-vdW and RPA@PBE are taken from Refs \citenum{shepard2019first} and \citenum{olsen2011dispersive}, respectively. Low-energy electron diffraction (LEED)\cite{gamo1997atomic} experiments measure the equilibrium separation of the graphene sheet on Ni(111) as $2.11\pm 0.07$ \r{A}, which is shown in Figure \ref{fig:Bivatiate_graphene_Ni}(b) as vertical dashed blue line with highlighted uncertainty. The corresponding binding energy has been measured by auger spectroscopy \cite{shelton1974equilibrium, janthon2013theoretical} as $74\pm8.1$ meV, illustrated by the horizontal dashed blue line with highlighted uncertainty.

Figure \ref{fig:Bivatiate_graphene_Ni}(b) shows that RPA@PBE predicts a chemisorption minimum with both the equilibrium position and the binding energy in good agreement with experimental data. Several studies and RPA@PBE support that graphene on Ni(111) has an additional minimum due to strong chemical and physical adsorption present in the graphene on Ni(111) system \cite{olsen2011dispersive, shelton1974equilibrium, janthon2013theoretical}. BEEF-vdW accurately predicts chemisorption separation distance, though the binding energy is greatly underestimated. Surprisingly, BEEF-vdW predicts a long-range physisorption minimum that is much deeper than its chemisorption minimum. Furthermore, RPBE+D3 shows a clear physisorption minimum at the expected separation, but the short-range chemisorption minimum is absent. Finally, the newly optimized Opt(MS+rVV10) functional agrees well by having a double minimum similar to RPA@PBE. Even though the binding energy predicted by the Opt(MS+rVV10) is slightly higher than the RPA@PBE and experimental uncertainties, the minimum intermolecular distance of chemisorption and physisorption along with the binding energies show the closest agreement to the benchmark and experimental data. {\color{blue}Even though the binding energy predicted by the Opt(MS+rVV10) is slightly higher than the RPA@PBE and experimental uncertainties, the minimum intermolecular distance of chemisorption and physisorption along with the binding energies show the closest agreement to the benchmark and experimental data. While our approach does not explicitly include many-body dispersion effects as implemented in advanced methods \cite{liu2012benzene,liu2014modeling}, the concurrent optimization of both the semilocal MS functional and rVV10 parameters may implicitly capture some collective effects. The MS2 meta-GGA's known ability to partially capture intermediate-range vdW interactions \cite{MS2_2}, combined with surface-specific parameterization, allows for some incorporation of the cooperative electronic response that characterizes many-body effects. In terms of computational efficiency, RPBE+D3 is the fastest among the tested vdW-corrected functionals. Opt(MS+rVV10) is approximately twice as expensive as RPBE+D3, whereas BEEF-vdW and SCAN+rVV10 are about four times more expensive. This trade-off between accuracy and computational cost may be an important consideration for large-scale or high-throughput studies.}

%{\color{blue} Additionally, the shape of the binding energy curve has the closest agreement to the RPA@PBE.} {\color{red} JWS: I don't think this is true. Did I miss something, or we didn't show the asymptotic here?} {\color{blue} Dr. Sun, I think this sentence is wrong, the binding energies of OPT(MS+rVV10) agree closely with the shape of the two minimums, however, they are not asymptotic.} 

\begin{figure*}
    \centering
    {\includegraphics[width=1\textwidth]{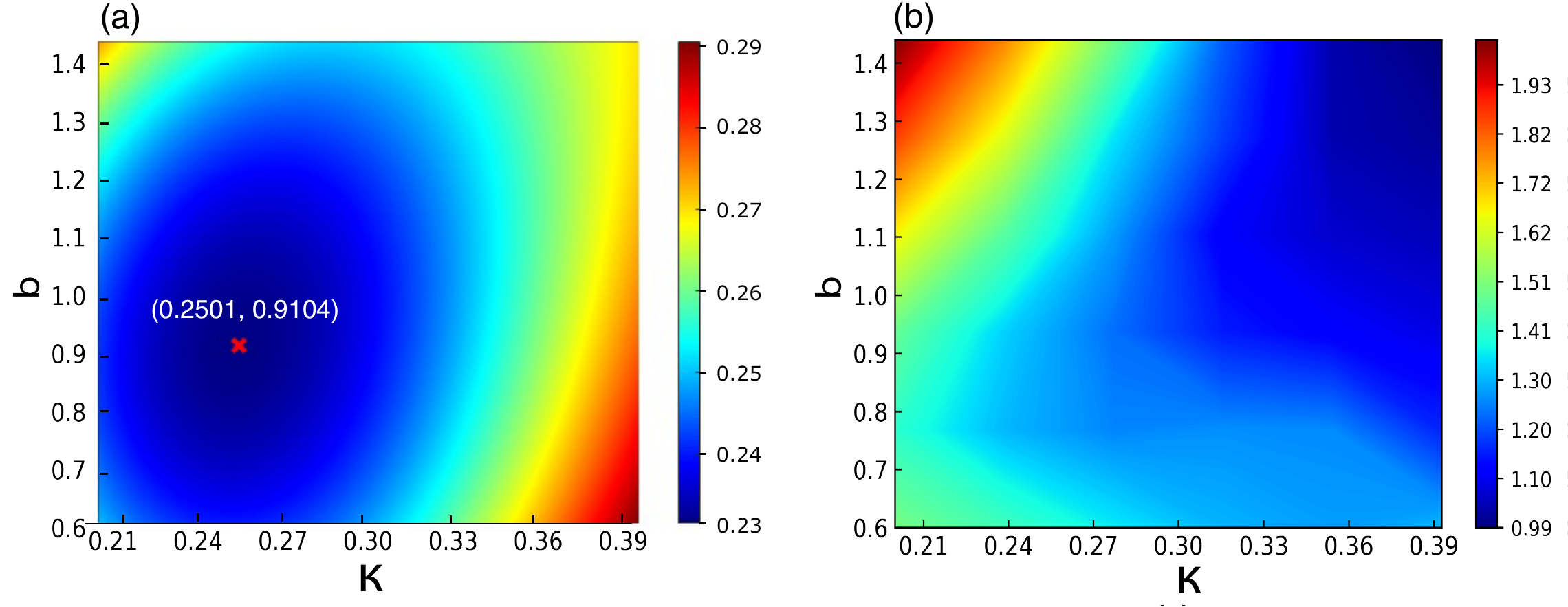}}
    \caption {(a) Heat map of 3$^{rd}$ order polynomial function fitted to Absolute Errors of CO adsorbed on Pt(111) surface as a function of $\kappa$ and $b$ fitting parameters, Errors are in eV. The minimizing parameters are marked in red as $\kappa = 0.2501$ and $b = 0.9104$. (b) Heat map of normalized absolute errors of atomization energies subset AE6\cite{lynch2003small}}  
    \label{fig:Heat_map_of_CO_Pt}
\end{figure*}

\subsection{Determination of parameters for O{pt}(MS+{r}VV10)} \label{sec_params}

We now present the procedure used to determine and optimize the parameters of the Opt(MS+rVV10) functional. The exchange-correlation energy of Opt(MS+rVV10) is expressed as:

\begin{equation}
E_\mathrm{xc}^\mathrm{opt(MS+rVV10)} = E_\mathrm{xc}^\mathrm{MS}[n; \kappa, b] + E_\mathrm{c}^{rVV10}[n; \beta]
\end{equation}

Here, $\kappa$ and $b$ are tunable parameters in the MS2 functional, while $\beta$ is a parameter in the rVV10 functional. In the original formulation of the MS2 meta-GGA,  $\kappa$ = 0.504 and $b$ = 4.0 were optimized to fit the atomization energies of six molecules (AE6 set), and the barrier heights of six reactions (BH6 set) \cite{lynch2003small}. Similarly in rVV10, $\beta$ = 6.3 was fitted to the interaction energies of 22 predominant dispersion-dominated complexes (S22 set) \cite{podeszwa2010improved}.

In this work, we re-optimized $\kappa$,  $b$, and $\beta$ to improve the functional's accuracy for both chemisorption and physisorption. Since the MS2 meta-GGA is known to partially capture intermediate-range vdW interactions \cite{sun2016accurate}, it was necessary to re-fit $\beta$ of rVV10 to avoid overestimating vdW forces, which could lead to double-counting. The $\beta$ parameter of rVV10 was therefore optimized for each ( $\kappa$, $b$ ) pair by fitting the binding energy curve of $Ar_{2}$, using highly accurate coupled-cluster CCSD(T) calculations as the reference \cite{patkowski2005accurate}. The $\kappa$ and $b$ of MS2 was then re-optimized by fitting to the experimental adsorption energy of CO on Pt(111), a well-studied surface system with reliable experimental data \cite{schiesser2010thermodynamics}.

We minimized the absolute error in the CO/Pt(111) adsorption energy with respect to $\kappa$ and b using a least-squares fit to a 3rd-order polynomial in the two-dimensional $\kappa$-$b$ space. This approach allowed us to efficiently explore the parameter space and identify the optimal values. 
% The error in the adsorption energy of CO/Pt(111) for each $\kappa$ and $b$ pair was fitted using a 3rd-order polynomial in two-dimensional $\kappa$-$b$ space. 
The resulting heat map (Figure \ref{fig:Heat_map_of_CO_Pt}(a)) revealed a well-defined minimum at $\kappa$=0.25 and $b$=0.91, with $\beta$ =26.26. This set of parameters delivered the lowest error for the CO/Pt(111) adsorption energy. 

When using AE6 atomization energies as the optimization objective for $\kappa$, and $b$, a very different minimum was found, located in the upper-right corner of the $\kappa$-$b$ parameter space (Figure \ref{fig:Heat_map_of_CO_Pt}(b)). A similar minimum emerged when both the AE6 and BH6 sets were used, in line with the original MS2 parameterization. These findings suggest that developing DFT exchange-correlation functionals for molecular adsorption on solid surfaces requires distinct optimization strategies compared to those used for general chemistry applications. {\color{blue}We acknowledge that the CE39 dataset is weighted toward chemisorbed systems, as noted in comprehensive benchmarking studies \cite{maurer2016adsorption}. However, our concurrent optimization approach against both CO/Pt(111) (chemisorption) and Ar$_2$ (physisorption) was specifically designed to address this imbalance and ensure balanced performance across both regimes, as evidenced by our bivariate plot results.
}

\noindent
\begin{table}[ht]
\footnotesize
\caption{Error statistics for the AE6, BH6, and LC20 test sets using RPBE+D3, BEEF-vdW, SCAN+rVV10 and Opt(MS+rVV10) dispersion corrected density functionals.} 
\centering
\vspace{2mm}
\begin{tabular}{crrrrrr}
\hline\hline
&RPBE+D3 & BEEF-vdW & SCAN+rVV10 & Opt(MS+rVV10) & \\ [0.2ex]
\hline
\multicolumn{6}{c}{Atomization energies (kcal/mol) of the AE6 molecules}\\ 
ME	& -6.0&	-2.8&	1.7&	-2.5\\ 
MAE	&8.2 &  4.3 &  3.8 & 5.8 \\ 
\multicolumn{6}{c}{Barrier Heights (kcal/mol) of the BH6 transition states} \\
ME &      -7.6	&-5.3&	-7.8&	-5.6   \\
MAE & 7.6 & 5.4 & 7.8 & 5.6 \\
\multicolumn{6}{c}{Lattice constants (\r{A}) of the LC20 solids}\\ 
ME&	0.070&	0.064&	-0.020&	0.008\\ 
MAE	& 0.075 & 0.071 & 0.021 & 0.020\\
\hline							
\hline
\end{tabular}
\label{table:subsets}
\end{table}

We conducted additional testing of the newly optimized Opt(MS+rVV10) functional on a set of small benchmark systems. These tests included the atomization energies from the AE6 set \cite{lynch2003small}, hydrogen transfer barrier heights from the BH6 set \cite{lynch2003small}, and lattice constants from the LC20 set \cite{sun2011self}, comprising 20 solids. The performance of Opt(MS+rVV10) was compared against other van der Waals-corrected functionals, including RPBE+D3, BEEF-vdW, and SCAN+rVV10 (the latter being a widely regarded general-purpose functional). 
The results are summarized in Table \ref{table:subsets}. For AE6 atomization energies, SCAN+rVV10 and BEEF-vdW showed better accuracy than Opt(MS+rVV10), which exhibited slightly higher errors. For BH6 barrier heights, both BEEF-vdW and Opt(MS+rVV10) performed well, with mean absolute errors (MAEs) of 5.4 kcal/mol and 5.6 kcal/mol, respectively. In the case of LC20 lattice constants, Opt(MS+rVV10) and SCAN+rVV10 were the most accurate, with MAEs of 0.020 Å and 0.021 Å, respectively. These results demonstrate that Opt(MS+rVV10) performs competitively in terms of accuracy, particularly for surface adsorption, while still maintaining reasonable accuracy in other general-purpose tasks.

\section{Conclusion}
We present the Opt(MS+rVV10), a novel exchange-correlation vdW-corrected functional, representing a significant advancement in the modeling of molecular adsorption on solid surfaces. This functional achieves a balanced performance in capturing both chemisorption and physisorption by concurrently optimizing its semilocal component and the long-range van der Waals correction. Internal parameters, including $\kappa$, $b$ in the MS2 functional and $\beta$ in rVV10, were re-fitted concurrently to the adsorption energy of CO/Pt(111) and the Ar$_2$ binding energy, respectively. The resulting Opt(MS+rVV10) functional demonstrates improved accuracy and computational efficiency compared to existing DFAs. Its rigorous parameterization demonstrates a significant improvement over other popular vdW-corrected functionals, such as BEEF-vdW, RPBE+D3, and SCAN+rVV10. This enhancement is evident from its reliable predictions across a broad spectrum of adsorption energies, as illustrated by its performance on the CE39 dataset.

Testing on graphene adsorption on Ni(111), a system known for its challenging double minima in the binding energy curve, revealed that Opt(MS+rVV10) closely aligns with the highly accurate RPA@PBE benchmark and experimental results. Furthermore, additional benchmarking on AE6, BH6, and LC20 sets showed that while Opt(MS+rVV10) offers slightly lower performance for AE6, it remains competitive for hydrogen transfer barriers and lattice constants, delivering overall reliable accuracy across diverse systems. The Opt(MS+rVV10) functional's improved accuracy in predicting both chemisorption and physisorption energies makes it particularly suited for studying complex catalytic systems such as zeolites, where both strong chemical bonding and weak van der Waals interactions play crucial roles in determining adsorption and reactivity of hydrocarbons. Additionally, its ability to capture subtle electronic interactions suggests potential applications in modeling transition metal catalysts for electrochemical CO$_2$ reduction, where an accurate description of adsorbate binding energies and charge transfer at the electrode-electrolyte interface is critical for predicting catalytic activity and selectivity \cite{o2016modelling}.

% This work paves the way for more accurate modeling of surface reactions, particularly in catalysis and materials science, where both chemical and physical interactions are essential. Future exploration could extend the applicability of Opt(MS+rVV10) to more complex systems, further enhancing its robustness and practical utility in surface science. 

\section{Computational Details}
The Opt(MS+rVV10) functional has been implemented in the developmental version of Vienna Ab-initio Simulation Package (VASP)\cite{Kresse94, Kresse96, Kresse99}. All calculations use the pseudopotential project-augmented wave method  \cite{blochl1994projector}, and a high-energy cutoff of 700 eV to truncate the plane wave basis set. Monkhorst-Pack \cite{Monkhorst}$k$ meshes for Brillouin zone integration are set to $6 \times 6 \times 1$ for $2 \times 2$ slabs, and $4 \times 4 \times 1$ for $3 \times 3$ slabs and $\Gamma$ point calculation for molecules. Ionic structures were relaxed to a residual force threshold of 0.01 eV/\r{A} per atom with a total energy tolerance of $10^{-5}$eV. 

Four-layer slabs with the bottom two layers fixed and 15 \r{A} vacuum between the surface species and the next repeated image in z-direction was used for adsorption energies. A dipole correction was applied on the adsorbate atom along the surface normal direction with magnitude dependent on charge and distance between the adsorbate and surface atom. Gas-phase molecules used in the calculation of the adsorption energies were optimized in a simulation cell of atleast 15 \r{A} vacuum. 

To calculate the binding curves for graphene  registry with the Ni(111) surface, we put the graphene sheet on top of the surface (as shown in Fig. 1 of Ref \cite{olsen2013random}). The metal surface was modeled with a four-layer slab generated with the experimental lattice constants and a 20 \r{A}-thick vacuum layer. The energy cutoff of 600 eV was used, and the $\Gamma$ centered $16\times 16 \times 1$ Monkhorst--Pack \cite{Monkhorst} $k$ meshes were used.

\section{Acknowledgement}
This work was supported by the U.S. DOE, Office of Science, Basic Energy Sciences (BES), Grant No. DE-SC0014208. M.K. also acknowledges the support of the American Chemical Society Petroleum Research Fund, grant number: PRF\# 59480-DNI5. Calculations were performed using the National Energy Research Scientific
Computing Center (NERSC) using NERSC Awards No. BES-ERCAP0020494 and No. BES-ERCAP0028709 and in part computational resources from Discovery, Research Computing at Northeastern University.  A.P. acknowledges support from Delaware Energy Institute, University of Delaware.  Q.Z. acknowledges support by Northeastern University, Chemical Engineering Department under start-up funding.

\section{Data and Code availability}
The patch code, data, main results, and the structures to reproduce this work are available at:   $\href{https://github.com/manishkothakonda/Opt-MS-rVV10-functional}{https://github.com/manishkothakonda/Opt-MS-rVV10-functional}$

\section{Notes}
The authors declare no competing financial interest.

\bibliography{achemso-demo}

\clearpage

\end{document}